\documentclass[twocolumn,english,prl,floatfix,showpacs]{revtex4-1}
\usepackage[T1]{fontenc}
\usepackage[latin9]{inputenc}
\usepackage{amsmath}
\usepackage{amssymb}
\usepackage{graphicx}
\usepackage{esint}

\makeatletter

\usepackage{babel}

\makeatother

\usepackage{babel}
\begin{document}

\title{Renormalization of the Unitary Evolution Equation for Coined Quantum
Walks}

\author{Stefan Boettcher$^{1}$, Shanshan Li$^{1}$, and Renato Portugal$^{2}$}

\affiliation{$^{1}$Department of Physics, Emory University, Atlanta, GA 30322;
USA ~\\
$^{2}$Laborat{ó}rio Nacional de Computa{ç}{ã}o Cient{í}fica,
Petropolis, RJ 25651-075; Brazil}
\begin{abstract}
We consider discrete-time evolution equations in which the stochastic
operator of a classical random walk is replaced by a unitary operator.
Such a problem has gained much attention as a framework for coined
quantum walks that are essential for attaining the Grover limit for
quantum search algorithms in physically realizable, low-dimensional
geometries. In particular, we analyze the exact real-space renormalization
group (RG) procedure recently introduced to study the scaling of quantum
walks on fractal networks. While this procedure, when implemented
numerically, was able to provide some deep insights into the relation
between classical and quantum walks, its analytic basis has remained
obscure. Our discussion here is laying the groundwork for a rigorous
implementation of the RG for this important class of transport and
algorithmic problems, although some instances remain unresolved. Specifically,
we find that the RG fixed-point analysis of the classical walk, which
typically focuses on the dominant Jacobian eigenvalue $\lambda_{1}$,
with walk dimension $d_{w}^{RW}=\log_{2}\lambda_{1}$, needs to be
extended to include the subdominant eigenvalue $\lambda_{2}$, such
that the dimension of the quantum walk obtains $d_{w}^{QW}=\log_{2}\sqrt{\lambda_{1}\lambda_{2}}$.
With that extension, we obtain analytically previously conjectured
results for $d_{w}^{QW}$ of Grover walks on all but one of the fractal
networks that have been considered.
\end{abstract}
\maketitle

\section{Introduction\label{sec:Intro}}

Quantum walks present one of the frameworks for which quantum computing
can satisfy its promise to provide a significant speed-up over classical
computation. Grover \cite{Gro97a} has shown that a quantum walk can
locate an entry in an unordered list of $N$ elements (i.e., sites
in some network) in a time that scales as $\sim\sqrt{N}$, a quadratic
speed-up over classical search algorithms. However, that finding was
based on a list in which all elements are interconnected with each
other, thus, raising the question regarding the impact of geometry
on this result. Note, for instance, that if the walk had to pass the
list over a linear, \emph{1d}-line of sites, no quantum effect would
provide an advantage over simply passing every site until the desired
entry is located. And for obvious engineering reasons, the design
of a quantum algorithm that could satisfy the Grover limit for lists
embedded in \emph{2d}-space is particularly desirable. Here, the issue
of geometry is especially pertinent, as it has been shown that only
discrete-time quantum walks with internal degrees of freedom (e.g.,
``coined'' walks) can attain the Grover-speedup for search on lattices
with $d\leq4$ \cite{Childs04,AKR05,Ambainis13,LiBo16}. Our discussion
here is focused on the long-time asymptotic properties of such coined
quantum walks.

While less dramatic as the super-polynomial speed-up of Shor's algorithm
for factoring \cite{Shor94}, searching is a far more common \cite{MM11}
and equally important task in the age of search engines \cite{Paparo13,chakraborty2015randomG}.
As fundamental as the random walk is to the description of randomized
algorithms in computer sciences \cite{MM11,Motwani95}, likewise quantum
walks have been established as a universal model of quantum computing
\cite{Childs09,Childs13}. Aside from their algorithmic relevance,
the \emph{physical} behavior of quantum walks, their entanglement,
localization, and interference effects \cite{Inui05,Falkner14a,QWNComms13,Boettcher14b,Ide2014}
in specific environments, rivals classical diffusion as an important
transport problem \cite{Weiss94,Redner01}. Already, there are numerous
experimental realizations of quantum walks, such as in wave-guides
\cite{Perets08,Martin11}, photonics \cite{Sansoni12,Crespi13}, and
optical lattices \cite{Weitenberg11}.

Whether we talk about random or about quantum walks, a complete description
for the spreading behavior is provided by the probability density
function (PDF) $\rho\left(\vec{x},t\right)$ to detect a walk at time
$t$ at site of distance $x=\left|\vec{x}\right|$ after starting
at the origin. For random walks at large times and spatial separations,
the PDF obeys the scaling collapse with the scaling variable $x/t^{1/d_{w}}$,
\begin{equation}
\rho\left(\vec{x},t\right)\sim t^{-\frac{d_{f}}{d_{w}}}f\left(x/t^{\frac{1}{d_{w}}}\right),\label{eq:collapse}
\end{equation}
where $d_{w}$ is the walk-dimension and $d_{f}$ is the fractal dimension
of the network \cite{Havlin87}. On a translationally invariant lattice
in any spatial dimension $d(=d_{f})$, it is easy to show that the
walk is always purely ``diffusive'', $d_{w}=2$, with a Gaussian
scaling function $f$. The scaling in Eq.~(\ref{eq:collapse}) still
holds when translational invariance is broken in certain ways or the
network is fractal (i.e., $d_{f}$ is non-integer). However, anomalous
diffusion with $d_{w}\not=2$ may arise in many of those transport
processes \cite{Havlin87,Bouchaud90,Weiss94,Hughes96}. This scaling
affects many important observables, for instance, the mean-square
displacement, $\left\langle x^{2}\right\rangle _{t}\sim t^{2/d_{w}}$,
or first-passage times \cite{Redner01,Condamin07,SWN}. 

We stipulate that a scaling relation like Eq.~(\ref{eq:collapse})
holds also for the coined quantum walk we are discussing in this paper.
For walks on regular lattices, Eq.~(\ref{eq:collapse}) indeed applies
in the ``weak limit'', but simply with $d_{w}^{QW}=1$ \cite{konno_2003a,grimmett_2004a}.
Beyond the regular lattice, the renormalization group (RG) is a good
method to explore the asymptotic scaling of a walk \cite{Havlin87,Hughes96,Redner01}.
Elucidating the effects of geometry and internal symmetries is exactly
the task that the real-space Renormalization group (RG) has been invented
for in the context of critical scaling in statistical physics \cite{Wilson71,Pathria}.
Indeed, the RG of classical random walks provides a straightforward
blueprint for developing the RG for a quantum walk, even with the
added complication of an internal coin space \cite{Boettcher13a}.
In this way, exact RG-flow equations for quantum walks on a number
of complex networks have been derived \cite{QWNComms13,Boettcher14b}.
Those results, for instance, have led to the conjecture that the walk
dimension $d_{w}$ in Eq.~(\ref{eq:collapse}) for a quantum walk
with a Grover coin always is \emph{half} of that for the corresponding
random walk, $d_{w}^{QW}=\frac{1}{2}\,d_{w}^{RW}$ \cite{Boettcher14b}. 

Unlike for the classical case, however, the analysis of the RG-flow
for quantum walks had only been conducted numerically, albeit with
high precision. The main obstacle for a rigorous treatment, and its
(partial) resolution we present here, is briefly stated as follows:
Note that the classical analysis of the RG-flow is determined by the
asymptotic behavior of \emph{real} poles in the complex-$z$ plane
after a Laplace-transformation from $t$ into $z$-space. For a quantum
walk, these Laplace-poles are \emph{complex} and behave in more subtle
ways. Those real poles merely flow radially in the complex-$z$ plane,
i.e., simply along the real-$z$ axis, impinging on the complex unit-circle
only at $z=1$, whereby the long-time behavior can be discerned \cite{Redner01}.
The corresponding complex poles in a quantum walk flow simultaneously
in radial and tangential directions, and impinge on the unit circle
in increasingly dense bands. The key observation in this paper is
that the corresponding radial flow of poles in the quantum walk, albeit
dominant, ultimately \emph{cancels} due to unitarity. As a consequence,
the otherwise sub-dominant tangential flow of poles actually controls
the walk dynamics. Based on this insight, we are able to derive some
\textendash{} but not all \textendash{} of the previously obtained
results analytically. The remaining obstacles indicate that certain
systems possess even more intricate scaling behavior than described
here.

In the following, we first highlight in Sec. \ref{sec:Unitary-Master-Equations}
the properties of a unitary evolution equation, its solution, and
its differences with a random walk equation. In Sec.\ \ref{sec:RG-for-QW},
we review the generic RG-evaluation of scaling in classical random
walks, followed by the corresponding derivation for quantum walks.
In Sec.\ \ref{subsec:Quantum-Walk:}, we apply the results of the
preceding RG-analysis to the specific cases of the known quantum walks on
fractals. We conclude with a discussion of the results in Sec.\ \ref{sec:Discussion}.

\section{Unitary Evolution Equations\label{sec:Unitary-Master-Equations}}

We consider the evolution equation for a classical or quantum walk
that is discrete in time and space,
\begin{equation}
\Psi\left(\vec{x},t+1\right)=\sum_{\vec{y}}{\cal U}\left(\vec{x},\vec{y}\right)\Psi\left(\vec{y},t\right),\label{eq:MasterE}
\end{equation}
where the propagator ${\cal U}\left(\vec{x},\vec{y}\right)$ is some
$M\times M$ matrix with $M$ reflecting a combination of a discrete
set of $N$ lattice sites and possibly a certain number of internal
degrees of freedom at each site. Assuming that we possess the eigensolution
for the propagator, ${\cal U}\phi_{j}=u_{j}\phi_{j}$ with an orthonormal
set of eigenvectors $\left\{ \phi_{j}\left(\vec{x}\right)\right\} _{j=1}^{M}$,
then the formal solution of Eq.~(\ref{eq:MasterE}) becomes
\begin{equation}
\Psi\left(\vec{x},t\right)=\sum_{\vec{y}}{\cal U}^{t}\left(\vec{x},\vec{y}\right)\Psi\left(\vec{y},0\right)=\sum_{j=1}^{M}a_{j}u_{j}^{t}\phi_{j}\left(\vec{x}\right).\label{eq:MasterExpansion}
\end{equation}

For a classical random walk, the site amplitude itself provides the
probability density, $\rho(\vec{x},t)=\Psi\left(\vec{x},t\right)$,
while for the quantum walk it is $\rho(\vec{x},t)=\left|\Psi\left(\vec{x},t\right)\right|{}^{2}$.
To preserve the norm $\sum_{\vec{x}}\rho(\vec{x},t)\equiv1$, the
propagator ${\cal U}$ is stochastic for a random walk, while it must
be \emph{unitary} for a quantum walk. For the stochastic operator
of a random walk, aside from the unique ($+1$)-eigenvalue of the
stationary state, the remaining eigenvalues have $\left|u_{j}\right|<1$,
thus, according to Eq.\ (\ref{eq:MasterExpansion}), the dynamics
is determined by $\rho(\vec{x},t)\sim e^{-\epsilon t}$ for large
times $t$ with $\epsilon=-\ln\max\left\{ \left|u_{j}\right|<1;1\leq j\leq M\right\} $.
In turn, for unitary ${\cal U}$ all eigenvalues are uni-modular,
$\left|u_{j}\right|=1$, such that $u_{j}=e^{i\theta_{j}}$ with real
$\theta_{j}$. Then, 
\begin{equation}
\rho\left(\vec{x},t\right)=r\left(\vec{x}\right)+\sum_{l<j}^{M}s_{j,l}\left(\vec{x}\right)\,\cos\left[\left(\theta_{j}-\theta_{l}\right)t\right],\label{eq:rho_t}
\end{equation}
where $r,s$ only depend on position and initial conditions. The cut-off
relevant for the long-time asymptotic behavior here is provided by
$\Delta\theta=\min\left\{ \left|\theta_{j}-\theta_{l}\right|>0;1\leq j,l\leq M\right\} $.
Furthermore, we note that a discrete Laplace transform (or generating
function), 
\begin{equation}
\overline{\Psi}\left(\vec{x},z\right)=\sum_{t=0}^{\infty}\Psi\left(\vec{x},t\right)\,z^{t},\label{eq:LaplaceTF}
\end{equation}
 of Eq.\ (\ref{eq:rho_t}) provides
\begin{equation}
\overline{\rho}\left(\vec{x},z\right)=\frac{C\left(\vec{x}\right)}{\prod_{j,l}^{M}\left[1-z\,e^{i\left(\theta_{j}-\theta_{l}\right)}\right]},\label{eq:rho_z}
\end{equation}
after placing all terms in the transformation of Eq.\ (\ref{eq:rho_t})
on their main denominator. Thus, all poles of $\overline{\rho}\left(\vec{x},z\right)$
in Eq.\ (\ref{eq:rho_z}), and hence for the site amplitudes $\overline{\Psi}\left(\vec{x},z\right)$,
are located right on the unit-circle in the complex-$z$ plane.

\section{Renormalization Group for Walks\label{sec:RG-for-QW}}

Instead of providing specific examples of walks on certain networks
here, we merely recount the essential details that allow us to efficiently
frame the RG analysis, its problem for quantum walks, and how we propose
to resolve it. An explicit derivation for the pedagogical case of
a walk on a 1$d$-line, for example, in parallel for the classical
and the quantum case, can be found in Ref. \cite{Boettcher13a}. 

To apply the renormalization group to a walk problem \cite{Redner01},
it is convenient to eliminate time $t$ in the evolution equation,
Eq.\ (\ref{eq:MasterE}), in the site basis via the discrete Laplace
transform in Eq.\ (\ref{eq:LaplaceTF}). Assuming a walk with an
initial condition at $t=0$ that is localized at the origin, we get
\begin{equation}
\overline{\Psi}\left(\vec{x},z\right)=\sum_{\vec{y}}z{\cal U}\left(\vec{x},\vec{y}\right)\overline{\Psi}\left(\vec{y},z\right)+\delta_{\vec{x},0}\psi_{IC}.\label{eq:masterLaplace}
\end{equation}
On exactly renormalizable networks, this linear system of equations
can now be decimated recursively by algebraically eliminating specific
site amplitudes such that the system remains self-similar after each
iteration \cite{QWNComms13,Boettcher14b}. Typically, this requires
the matrix $z{\cal U}$ to be sparse and its non-zero coefficients
should be re-presentable in terms of just a small set $\vec{a}_{0}(z)=\left(a_{0},b_{0},c_{0},\ldots\right)$
of site-independent ``hopping parameters'' of the unrenormalized
state. Each iteration constitutes an RG-step from level $k$ to $k+1$,
starting at the unrenormalized state with $k=0$. Each RG-step represents
a coarse-graining of the system, such that each remaining amplitude
at level $k$ represents a spatial domain of length $L=b^{k}$, and
the $\vec{a}_{k}$ the effective transitions between those, where
$b$ is the rescaling of length in each step. (It is $b=2$ in most
cases referred to here.) These ``renormalized'' $\vec{a}_{k}$ arise,
as the remaining equations attain self-similarity only for an appropriate
redefinition of the hopping parameters, leading to the mapping
\begin{equation}
\vec{a}_{k+1}\left(z\right)={\cal RG}\left[\vec{a}_{k}\left(z\right)\right].\label{eq:RGflow}
\end{equation}
which is called the RG-flow. This set of coupled, rational maps exactly
encapsulates the entire walk process. 

While the RG-flow in Eq.\ (\ref{eq:RGflow}) usually can not be solved
in general, the properties of its fixed point(s)
\begin{equation}
\vec{a}_{\infty}={\cal RG}\left(\vec{a}_{\infty}\right)\label{eq:FP}
\end{equation}
for $k\to\infty$ can be explored to reveal the dynamics of the walk
asymptotically at large length- and time-scales. Specifically, $k\to\infty$
corresponds to a diverging system size $N=L^{d_{f}}$ while $\left|z\right|\to1$
according to Eq.\ (\ref{eq:LaplaceTF}) accesses the large-$t$ limit,
as the hopping parameters $\vec{a}_{k}(z)$ become ever-more complicated
rational functions in $z$ under the RG-flow in Eq.\ (\ref{eq:RGflow}).
We can linearize the RG-flow via the Ansatz
\begin{equation}
\vec{a}_{k}\left(z\right)\sim\vec{a}_{\infty}+(1-z)\vec{\alpha}_{k},\label{ak_alpha}
\end{equation}
for $z\to1$ and $k\to\infty$, assuming $(1-z)\vec{\alpha}_{k}\ll1$.
Then, we get the linear system
\begin{equation}
\vec{\alpha}_{k+1}=\vec{\alpha}_{k}\circ J,\label{eq:linearizedRG}
\end{equation}
with the Jacobian matrix
\begin{equation}
J=\left(\left.\frac{\partial{\cal RG}}{\partial\vec{a}_{k}}\right|_{k\to\infty}\right),\label{eq:Jacobian}
\end{equation}
such that the solutions of Eq.\ (\ref{eq:linearizedRG}) are linear
combinations,
\begin{equation}
\vec{\alpha}_{k}\sim\lambda_{1}^{k}\vec{v}_{1}{\cal A}_{1}+\lambda_{2}^{k}\vec{v}_{2}{\cal A}_{2}+\ldots,\label{eq:alphaksolution}
\end{equation}
where $\lambda_{j}$ are the eigenvalues of $J$ in descending order,
and $\vec{v}_{j}$ the associated eigenvectors. (Since $J$ is not
necessarily Hermitian, the eigenvectors are not necessarily orthogonal!)

\begin{figure}
\hfill{}\includegraphics[bb=70bp 355bp 520bp 560bp,clip,width=1\columnwidth]{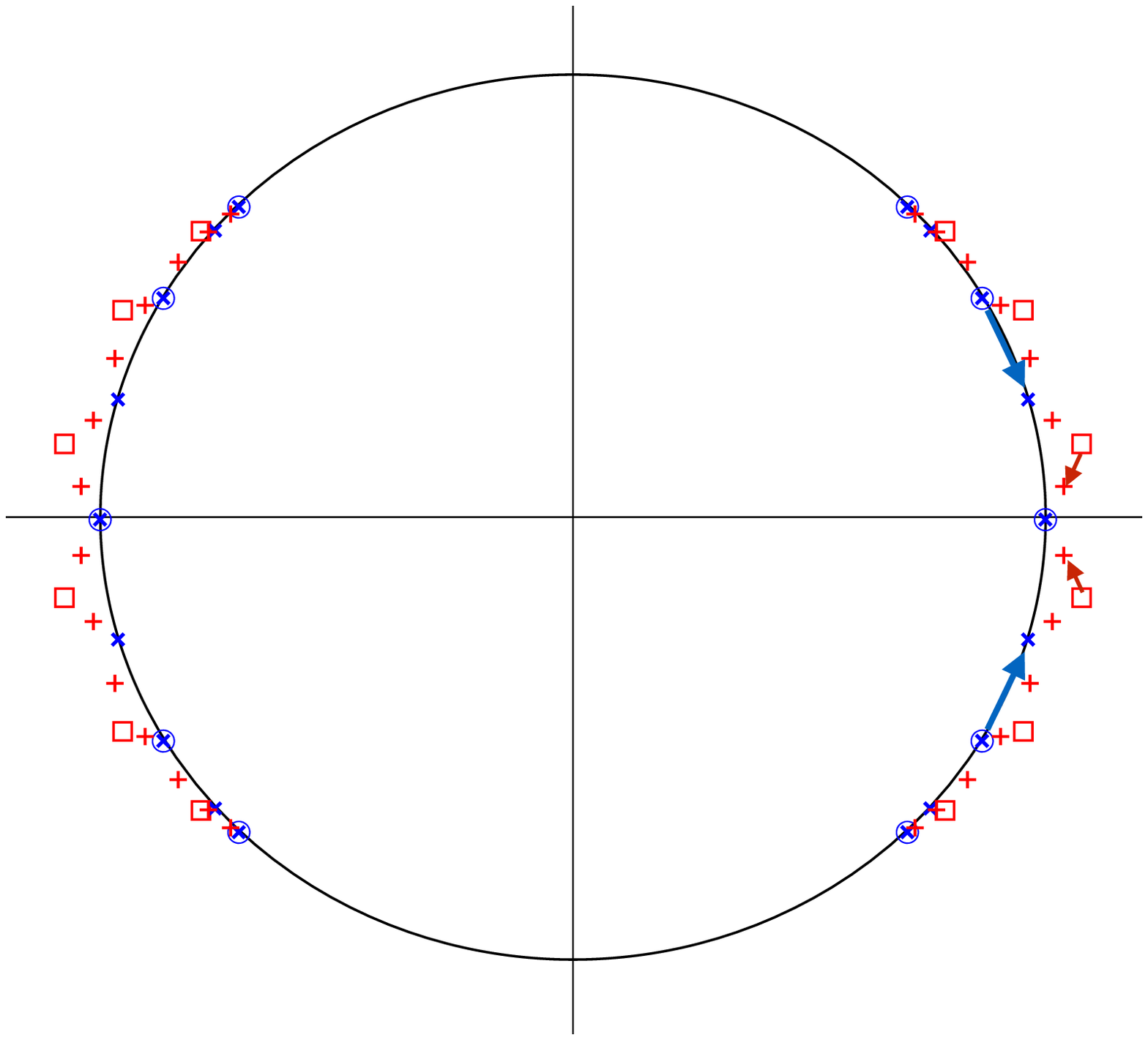}\hfill{}

\medskip{}

\hfill{}\includegraphics[bb=120bp 335bp 610bp 545bp,clip,width=1\columnwidth]{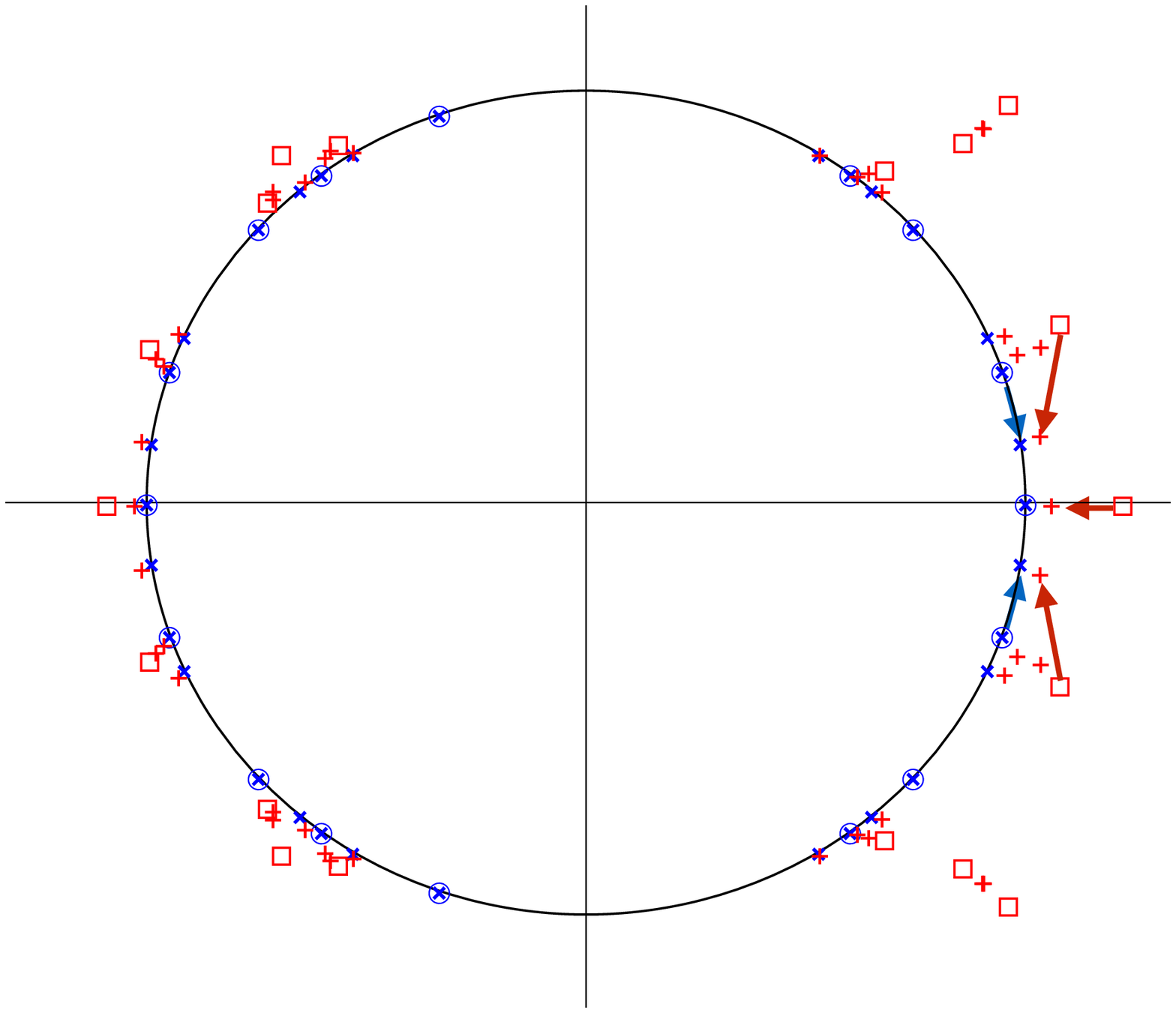}\hfill{}\caption{\label{fig:DSGpoles}Plot of the poles of the Laplace transforms in
the complex-$z$ plane at RG-steps $k=2$ and $k=3$ for some site-amplitude
$\overline{\psi}_{0}^{(k)}(z)$ (in blue, with $\circ$ for $k=2$
and $\times$ for $k=3$) and for the hopping parameters $\vec{a}_{k}(z)$
(in red, with $\square$ for $k=2$ and $+$ for $k=3$) for quantum
walk on the 1$d$-line (top) and the dual Sierpinski gasket (bottom).
(In these walks, poles are certain to occur in complex-conjugate pairs,
so only the upper $z$-plane is shown.) Even at such small orders
$k$, there is a large number of poles, which proliferate exponentially
with $k$. All poles for the site-amplitude are on the unit-circle,
while all poles for the hopping parameters are outside. Although the
pattern by which poles evolve appears complicated, either type of
poles impinge increasingly on the real-$z$ axis. Especially marked
is the RG-flow of the poles closest to $z=1$, both for the site-amplitude
(blue arrows) and the hopping parameters (red arrows). While the former
only flow tangentially along the circle, the later displace tangentially
as well as radially, with one pole potentially flowing just radially
inward along the real axis (bottom panel).}
\end{figure}

As Sec.\ \ref{sec:Unitary-Master-Equations} suggests, especially
the Laplace-poles closest to $\left|z\right|\to1$ assume an important
role. Now, the location of the poles in the complex-$z$ plane for
any observable, like the PDF $\overline{\rho}(\vec{x},z)$, do not
necessarily correspond to those of $\vec{a}_{k}\left(z\right)$, although
$\overline{\rho}(\vec{x},z)=f_{\vec{x}}\left[\vec{a}_{k}\left(z\right)\right]$
is a \emph{functional} of the hopping parameters. However, this distinction
does not pose a problem for the real poles in the classical walk:
either set of poles flows radially along the real-$z$ axis towards
$z=1$, exhibiting the same scaling. In contrast, quantum-walk observables
are unitary and therefore have strictly uni-modular poles, such as
in Eq.\ (\ref{eq:rho_z}), that flow only tangentially on the unit
circle in the complex-$z$ plane, while the renormalized hopping parameters
$\vec{a}_{k}\left(z\right)$ individually have poles that flow radially
\emph{and} tangentially, with no obvious connection a-priori, see
Fig.\ \ref{fig:DSGpoles}. In the following, we explore this distinction. 

\subsection{Classical Fixed-Point Analysis\label{subsec:Classical-Fixed-Point-Analysis}}

For a classical random walk, almost all poles of $\vec{a}_{k}\left(z\right)$
are on the real-$z$ axis with $z>1$. Let $z_{k}$ be the pole that
is closest to $z=1$, to wit, 
\begin{equation}
z_{k}\sim1+\epsilon_{k},\qquad\left(\epsilon_{k}\to0\right)\label{eq:zk}
\end{equation}
with real $\epsilon_{k}>0$. Then, the generic form for a simple pole
near $z=1$, with $\vec{a}_{k}\left(1\right)=\vec{a}_{\infty}$ at
the fixed point in Eq.\ (\ref{eq:FP}), is:
\begin{eqnarray}
\vec{a}_{k}\left(z\right) & \sim & \frac{1-z_{k}}{z-z_{k}}\,\vec{a}_{\infty}\sim\vec{a}_{\infty}\left[1-\frac{1}{\epsilon_{k}}\left(1-z\right)+\ldots\right],\label{eq:ak_exp}
\end{eqnarray}
which is the expected behavior for the hopping parameters that justifies
the Ansatz in Eq.\ (\ref{ak_alpha}). Note that $1/\epsilon_{k}$
is not only the most divergent term of order $1-z$, it is the only
divergence possible. This is reflected in Eq.\ (\ref{eq:alphaksolution})
in the fact that there is only a single divergent eigenvalue, $\lambda_{1}>1$,
of the Jacobian in Eq.\ (\ref{eq:Jacobian}) for a classical walk.
The comparison implies 
\begin{equation}
\epsilon_{k}\sim\left|\vec{\alpha}_{k}\right|^{-1}\sim\lambda_{1}^{-k}\label{eq:alphaeps}
\end{equation}
at the cross-over $(1-z)\alpha_{k}\sim1$ that determines the cut-off. 

With $\overline{\rho}(\vec{x},z)=f_{\vec{x}}\left[\vec{a}_{k}\left(z\right)\right]$,
the classical PDF now attains the form for $z\to1$:
\begin{equation}
\overline{\rho}(\vec{x},z)\sim\frac{1-z_{k}^{\prime}}{z-z_{k}^{\prime}}\,A,\label{eq:psik}
\end{equation}
where its closest pole $z_{k}^{\prime}\sim1+C\epsilon_{k}$ may differ
from $z_{k}$, but only by a constant $C$ that does not affect the
scaling. To see this, we consider the inverse Laplace transform and
inserting Eq.\ (\ref{eq:psik}):
\begin{equation}
\rho\left(\vec{x},t\right)=\ointop\frac{dz}{2\pi iz}\,z^{-t}\overline{\rho}(\vec{x},z)\sim AC\epsilon_{k}e^{-C\epsilon_{k}t},
\end{equation}
where $\epsilon_{k}$ provides the cut-off. If we now calculate temporal
moments, say, of the first passage time at some site $\vec{x}$ \cite{Redner01},
it is
\begin{eqnarray}
\left\langle t^{n}\right\rangle _{k} & = & \frac{1}{{\cal N}}\int_{0}^{\infty}dt\,t^{n}\rho(\vec{x},t)\sim\epsilon_{k}^{-n},\label{eq:classicalMoment}
\end{eqnarray}
where the norm ${\cal N}$ absorbs any factor so that $\left\langle t^{0}\right\rangle _{k}\equiv1$.
Then, Eq.\ (\ref{eq:alphaeps}) implies for the scaling of the characteristic
time-scale $T$ of the dynamics associated with system size $N$ that
\begin{equation}
T=\left[\left\langle t^{n}\right\rangle _{k}\right]^{\frac{1}{n}}\sim\epsilon_{k}^{-1}\sim\lambda_{1}^{k}\sim\left(2^{k}\right)^{\log_{2}\lambda_{1}}\sim L^{d_{w}}\sim N^{\frac{d_{w}}{d_{f}}},\label{eq:normmoment}
\end{equation}
such that the classical walk dimension becomes
\begin{equation}
d_{w}^{RW}=\log_{2}\lambda_{1}.\label{eq:dw}
\end{equation}

\subsection{Fixed-Point Analysis of the Quantum Walk\label{subsec:Quantum-Walk:}}

For the quantum walk, Fig.\ \ref{fig:DSGpoles} suggests that the
poles of the hopping parameters reside in certain bands near but not
on the unit circle of the complex-$z$ plane. (All poles come in complex-conjugate
pairs for a unitary quantum walk with a purely real coin, such as
the Grover coin \cite{PortugalBook}.) As for the classical walk,
we shall assume that poles closest to $z=\pm1$, i.e., those on or
near the real-$z$ axis, are again the most relevant. This assumption
will prove appropriate but more difficult to justify than in the classical
case, as we will discuss below. The key observation now is that the
flow of these poles in radial and tangential directions can be parametrized
as 
\begin{equation}
z_{k}^{(\pm)}\sim\left(1+\epsilon_{k}\right)e^{\pm i\theta_{k}}\sim1\pm i\theta_{k}+\epsilon_{k},\quad\left(1\gg\theta_{k}\gg\epsilon_{k}\gg\theta_{k}^{2}\right)\label{eq:zkpm}
\end{equation}
for large order $k$ in the RG-flow, Eq.\ (\ref{eq:RGflow}), while
there also might be a real pole, $z_{k}^{(0)}\sim1+\epsilon_{k}$.
Previous discussions \cite{QWNComms13,Boettcher14b} suggest that
radial flow is faster than tangential flow ($\theta_{k}\gg\epsilon_{k}$)
for these poles. This follows from the scaling collapse found in Refs.
\cite{QWNComms13,Boettcher14b} that lines up diverging features (i.e.,
poles close to the unit circle) in the tangential ($\theta_{k}$)
direction, while their width (corresponding to the radial proximity
to the circle, measured by $\epsilon_{k}$) declines more rapidly.

Then, like in Eq.\ (\ref{eq:ak_exp}) for the classical case, the
most generic expression for any hopping parameter is: 
\begin{align}
\vec{a}_{k}(z) & \sim\vec{a}_{\infty}\left[\frac{pe^{i\phi}}{2\cos\phi}\,\frac{1-z_{k}^{(+)}}{z-z_{k}^{(+)}}+\frac{pe^{-i\phi}}{2\cos\phi}\,\frac{1-z_{k}^{(-)}}{z-z_{k}^{(-)}}\right.\label{eq:psiQW}\nonumber\\
 & \qquad\qquad\left.+(1-p)\,\frac{1-z_{k}^{(0)}}{z-z_{k}^{(0)}}\right]
\end{align}
with possibly ($p<1$) a real pole at $z_{k}^{(0)}$ and complex conjugate
poles at $z_{k}^{(\pm)}=\left[z_{k}^{(\mp)}\right]^{*}$. Note that
again $\vec{a}_{k}(1)=\vec{a}_{\infty}$ at the fixed point, but that
we also assume $\vec{a}_{k}\left(z^{*}\right)=\left[\vec{a}_{k}(z)\right]^{*}$.
Here, $p$ determines the balance in weight between real and complex
roots, while $\phi\not=0$ allows for a complex residue at the pole.
Expanding similar to Eq.\ (\ref{eq:ak_exp}) for the classical case,
we find 
\begin{equation}
\vec{a}_{k}\left(z\right)\sim\vec{a}_{\infty}\left(1-\left[\frac{r}{\epsilon_{k}}+s\,\begin{cases}
\frac{\epsilon_{k}}{\theta_{k}^{2}}, & \phi=0,\\
\frac{1}{\theta_{k}}, & \phi\not=0
\end{cases}\right]\left(1-z\right)+\ldots\right),\label{eq:ak_QM}
\end{equation}
with some constants $r,s$. Thus, allowing for the tangential flow
in Eq.\ (\ref{eq:zkpm}) produces a second independent divergence
at order $1-z$ in Eq.\ (\ref{eq:ak_QM}). Hence, we expect both,
leading and sub-leading eigenvalues $\lambda_{1,2}>1$ in Eq.\ (\ref{eq:alphaksolution})
for the respective expansion near the fixed point, to match up with
Eq.\ (\ref{eq:ak_QM}). As in the classical case, Eq.\ (\ref{eq:alphaeps}),
it is typically $\epsilon_{k}\sim\lambda_{1}^{-k}$ as the most-divergent
contribution in Eq.\ (\ref{eq:ak_QM}), although interchanging the
roles of $\lambda_{1,2}$ does not affect the following outcome: Comparing
Eq.\ (\ref{eq:ak_QM}) with Eq.\ (\ref{eq:alphaksolution}) assuming
real residues ($\phi=0$), it is $\epsilon_{k}/\theta_{k}^{2}\sim\lambda_{2}^{k}$,
i.e., 
\begin{equation}
\theta_{k}\sim\left(\sqrt{\lambda_{1}\lambda_{2}}\right)^{-k}.\label{eq:theta12}
\end{equation}

\section{Discussion\label{sec:Discussion}}

While it would seem futile to consider a subdominant contribution
to scaling, we have shown that, due to unitarity, it actually becomes
the key to understand the long-range dynamics of a quantum walk. We
know that there is a functional relation between the hopping parameters
and the PDF, $\overline{\rho}(\vec{x},z)=f_{\vec{x}}\left[\vec{a}_{k}\left(z\right)\right]$,
and we are \emph{assured} by Eq.\ (\ref{eq:rho_z}) that all the
poles of the PDF are (products of) uni-modular modes. Therefore, the
functional must be such that the leading scaling resulting from the
radial flow of poles in $\vec{a}_{k}\left(z\right)$ will cancel in
the PDF. This fact, as exhibited in Fig.\ \ref{fig:DSGpoles}, we
show explicitly for the case of the quantum walk on a line in the
Appendix below. Again, it is guaranteed to occur due to Eq.\ (\ref{eq:rho_z})
but a general proof would be useful. 

As Eq.\ (\ref{eq:rho_z}) further suggests, however, it would appear
that it is not sufficient to merely focus on those poles closest to
the real-$z$ axis with the smallest $\theta_{k}$. After all, it
is the smallest \emph{difference} $\Delta\theta$ in the arg of two
poles that sets the cut-off. But we will now argue that there should
be only \emph{one} scale, $\theta_{k}$ as in Eq.\ (\ref{eq:theta12}),
for the tangential flow of poles: Poles anywhere along the unit circle
either (1) scale with $\theta_{k}$, like those closest to the real
axis, or (2) may converge to some complex constant, in which case
their correction scales like $\theta_{k}$. Under that assumption,
$\Delta\theta$ is either the difference of two type-(1) poles, two
type-(2) poles, or a type-(1) and a type-(2) pole. The first case
again leads to $\theta_{k}$ itself again, the second case either
leads to another constant or, if the two leading constants of the
type-(2) poles cancel, it leads back to the first case, and the third
case merely converges to the constant part of the type-(2) pole. Consequently,
we find a minimal $\Delta\theta\sim\theta_{k}$, even if the poles
that contribute to the smallest difference are not those closest to
the real-$z$ axis and the fixed point there. Yet, observing these
poles \emph{does} provide the relevant scaling, $\theta_{k}$ in Eq.\ (\ref{eq:theta12}).

To justify these conclusions, we can re-assess some of the previous
results and the conjecture of $d_{w}^{QW}=\frac{1}{2}\,d_{w}^{RW}$
\cite{QWNComms13,Boettcher14b}. The easiest case of quantum walk on
a 1$d$-line, in which all hopping parameters and any observable can
be calculated not only asymptotically but also in closed form, unfortunately
does not provide much insight here: The RG analysis \cite{Boettcher13a}
yields degenerate Jacobian eigenvalues, $\lambda_{1}=\lambda_{2}=2$,
such that there is no distinction, i.e., $\lambda_{1}\equiv\sqrt{\lambda_{1}\lambda_{2}}$,
and one would be lead to the false conclusion that a naive classical
analysis as in Sec.\ \ref{subsec:Classical-Fixed-Point-Analysis}
would suffice. In this case, we can merely ascertain (in the Appendix)
that the tangential flow of poles in the hopping parameters translates
exactly into that of the poles belonging to observable while the radial
flow cancels. Although this can not be shown generally but only for
small system sizes in the non-trivial fractal networks, these in turn
allow to validate Eq.\ (\ref{eq:theta12}): The RG for quantum walk on
the dual Sierpinski gasket (DSG) \cite{QWNComms13} yields $\lambda_{1}=3$,
$\lambda_{2}=\frac{5}{3}$, and $\lambda_{3}=1$, such that $\sqrt{\lambda_{1}\lambda_{2}}=\sqrt{5}$
provides the numerically determined scaling, $d_{w}^{QW}=\log_{2}\sqrt{5}$,
and $\lambda_{3}\not>1$ remains irrelevant. Since classically $\lambda_{1}^{{\rm RW}}=5$,
it verifies the conjecture of $d_{w}^{QW}=\frac{1}{2}\,d_{w}^{RW}$.
The same pattern holds for the two Migdal-Kadanoff networks analyzed
in Ref. \cite{Boettcher14b}: For the 3-regular network called MK3,
the Jacobian eigenvalues are $\lambda_{1}=7$ and $\lambda_{2}=3$
such that $\lambda_{1}\lambda_{2}=\lambda_{1}^{{\rm RW}}=21$, and
for the 4-regular network called MK4 they are $\lambda_{1}=13$ and
$\lambda_{2}=\frac{19}{7}$, such that $\lambda_{1}\lambda_{2}=\lambda_{1}^{{\rm RW}}=\frac{247}{7}$.
It is surprising, then, that for the 3-regular Hanoi network called
HN3 \cite{SWPRL}, Ref. \cite{Boettcher14b} found $\lambda_{1}=2$
and $\lambda_{2}=\left(1+\sqrt{17}\right)/4$, so that $\lambda_{1}\lambda_{2}\not=\lambda_{1}^{{\rm RW}}=2\left(\sqrt{5}-1\right)$
is unrelated to the classical eigenvalue, yet, numerically the conjecture
still appears to hold. This suggests that the relation in Eq.\ (\ref{eq:theta12})
is not quite so fundamental, i.e., some of the assumptions leading
to it are violated by HN3. Not even the simple alternative of complex
residues at the poles, $\phi\not=0$ in Eq.\ (\ref{eq:ak_QM}), implying
$\theta_{k}\sim\lambda_{2}^{-k}$, can explain the discrepancy. Furthermore,
it appears that a more general argument than Eq.\ (\ref{eq:theta12})
must exists to justify the conjecture $d_{w}^{{\rm QW}}=\frac{1}{2}d_{w}^{{\rm RW}}$.
Finally, we note in passing that in all cases of quantum walks, the
dominant Jacobian eigenvalue happens to provide the fractal exponent
of the network itself, $\log_{2}\lambda_{1}=d_{f}$.

\paragraph*{Acknowledgements:}

SB and SL acknowledge financial support from the U. S. National Science
Foundation through grant DMR-1207431. SB acknowledges financial support
from CNPq through the ``Ciência sem Fronteiras'' program. SL acknowledges
financial support from the American Physical Society through the Brazil-U.S.
Exchange Program. Both, SB and SL thank LNCC for its hospitality.
RP acknowledges financial support from Faperj and CNPq.

\bibliographystyle{apsrev4-1}
\bibliography{/Users/stb/Boettcher}

\section*{Appendix\label{sec:Appendix}}

\subsection*{Poles of Hopping Parameters and Observables\label{subsec:Poles-of-Hopping}}

Here we show that poles in Laplace-space for the hopping parameters
can not be on the unit circle of the complex-$z$ plane, while in
turn the poles of the site-amplitudes are \emph{only} on that circle,
as shown in Fig.\ \ref{fig:DSGpoles}. While this holds for any system
due to Eq.\ (\ref{eq:rho_z}), we demonstrate this here by example
of a quantum walk on a 1$d$-line with periodic boundary conditions. 

The propagator ${\cal U}$ on the 1\emph{d}-line is given by
\begin{eqnarray}
{\cal U} & = & \sum_{n}\left\{ A\,\delta_{n,n+1}+B\,\delta_{n,n-1}+M\,\delta_{n,n}\right\} ,\label{eq:propagator}
\end{eqnarray}
where the general form of the hopping operators $A$, $B$, and $M$
is only constrained by the requirement of unitarity:
\begin{eqnarray}
{\cal I} & = & {\cal U}^{\dagger}{\cal U},\nonumber \\
 & = & \sum_{n}\sum_{m}\left\{ A^{\dagger}\delta_{n,n+1}+B^{\dagger}\delta_{n,n-1}+M^{\dagger}\delta_{n,n}\right\} \nonumber \\
 &  & \qquad\left\{ A\delta_{m,m+1}+B\delta_{m,m-1}+M\delta_{m,m}\right\} ,\label{eq:unitaryU}\\
 & = & \sum_{n}\left\{ \left(A^{\dagger}A+B^{\dagger}B+M^{\dagger}M\right)\delta_{n,n}+\left(A^{\dagger}M+M^{\dagger}B\right)\delta_{n,n+1}\right.\nonumber \\
 &  & \quad\left.+\left(B^{\dagger}M+M^{\dagger}A\right)\delta_{n,n-1}+A^{\dagger}B\delta_{n,n+2}+B^{\dagger}A\delta_{n,n-2}\right\} ,\nonumber 
\end{eqnarray}
which is satisfied by the hopping matrices:
\begin{eqnarray}
{\cal I} & = & A^{\dagger}A+B^{\dagger}B+M^{\dagger}M,\nonumber \\
0 & = & A^{\dagger}M+M^{\dagger}B=\left(B^{\dagger}M+M^{\dagger}A\right)^{\dagger},\nonumber \\
0 & = & A^{\dagger}B=\left(B^{\dagger}A\right)^{\dagger}.\label{eq:unitarityCond}
\end{eqnarray}
Adding these relations implies that the sum $A+B+M$ itself is unitary.
Due to the self-similarity of any renormalized structure, this property
will remain true after each RG-step $k$. For example, from Eq.\ (\ref{eq:masterLaplace})
applied to Eq.\ (\ref{eq:propagator}), we learn that in Laplace-space
the hopping parameters at the initial ($k=0$) RG-step have the from
$A_{0}=zA$, $B_{0}=zB$, and $M_{0}=zM$ and subsequently become
complicated algebraic expressions of $z$ via the RG-flow in Eq.\ (\ref{eq:RGflow}).
Then, while on the unit circle in the complex-$z$ plane, we have
at any $k$ that
\begin{equation}
\det\left[A_{k}\left(z\right)+B_{k}\left(z\right)+M_{k}\left(z\right)\right]=e^{i\xi_{k}\left(z\right)}\qquad\left(\left|z\right|=1\right)\label{eq:det_unitary}
\end{equation}
with some real phase $\xi_{k}\left(z\right)$.

\subsubsection*{Poles of the renormalized hopping paramters}

While the conditions in Eq.\ (\ref{eq:unitarityCond}) are more generally
valid for any \emph{r}-dimensional matrices with $r\geq2$, we restrict
ourselves to the simplest case $r=2$. The renormalization group treatment
of the quantum walk on the 1$d$-line \cite{Boettcher13a} entails
a decomposition of the hopping matrices into the unitary $2\times2$
coin matrix ${\cal C}$ and $k$-th renormalized shift matrices 
\begin{equation}
P_{k}=\left(\begin{array}{cc}
a_{k} & 0\\
0 & 0
\end{array}\right),\quad Q_{k}=\left(\begin{array}{cc}
0 & 0\\
0 & -a_{k}
\end{array}\right),\quad R_{k}=\left(\begin{array}{cc}
0 & b_{k}\\
b_{k} & 0
\end{array}\right),\label{eq:PQR}
\end{equation}
such that $A_{k}=P_{k}{\cal C}$, $B_{k}=Q_{k}{\cal C}$, and $M_{k}=R_{k}{\cal C}$.
Here, $a_{k}\left(z\right)$ and $b_{k}\left(z\right)$ are the scalar
renormalized hopping parameters that obey an RG-flow as described
in Eq.\ (\ref{eq:RGflow}). (For the explicit form, of these recursions,
see Ref. \cite{Boettcher13a}.) Since the coin ${\cal C}$ is unitary,
the shift matrices then must satisfy the same conditions in Eq.\ (\ref{eq:unitarityCond}).
While the last two relations are satisfied automatically for matrices
in the form of Eq.\ (\ref{eq:PQR}), the first relation implies that
\begin{equation}
\left|a_{k}\left(z\right)\right|^{2}+\left|b_{k}\left(z\right)\right|^{2}=1\qquad\left(\left|z\right|=1\right).\label{eq:RGnorm}
\end{equation}
This proves that neither $a_{k}$ nor $b_{k}$ can have any singularities
on the unit circle $\left|z\right|=1$.

We finally remark that Eq.\ (\ref{eq:RGnorm}) in itself does not
imply that there is a constraint between $a_{k}$ and $b_{k}$ that
could make one of these redundant. Eq.\ (\ref{eq:RGnorm}) fixes
their relation only up to a phase that could itself depend on $k$
and $z$. For example, the alternative unitarity condition derived
from Eq.\ (\ref{eq:det_unitary}) would say:
\begin{equation}
a_{k}^{2}\left(z\right)+b_{k}^{2}\left(z\right)=e^{i\xi_{k}\left(z\right)}.
\end{equation}

\subsubsection*{Poles of the site amplitudes}

Although we have already shown in Eq.\ (\ref{eq:rho_z}) on very
general grounds that the poles of the site-amplitudes are all located
on the unit-circle in the complex-$z$ plane, it is quite instructive
to see how this fact emerges in each particular case, especially in
relation with the hopping parameters. In our example of the 1$d$-line
of $N=2^{k}$ sites with periodic boundary conditions, the loop merely
consists of four remaining sites after $k-2$ RG-steps, with
\begin{eqnarray}
\overline{\psi}_{0} & = & A_{k-2}\overline{\psi}_{\frac{N}{4}}+B_{k-2}\overline{\psi}_{\frac{3N}{4}}+M_{k-2}\overline{\psi}_{0}+\psi_{IC},\nonumber \\
\overline{\psi}_{\frac{N}{4}} & = & A_{k-2}\overline{\psi}_{\frac{N}{2}}+B_{k-2}\overline{\psi}_{0}+M_{k-2}\overline{\psi}_{\frac{N}{4}},\nonumber \\
\overline{\psi}_{\frac{N}{2}} & = & A_{k-2}\overline{\psi}_{\frac{3N}{4}}+B_{k-2}\overline{\psi}_{\frac{N}{4}}+M_{k-2}\overline{\psi}_{\frac{N}{2}},\label{eq:loop4search}\\
\overline{\psi}_{\frac{3N}{4}} & = & A_{k-2}\overline{\psi}_{0}+B_{k-2}\overline{\psi}_{\frac{N}{2}}+M_{k-2}\overline{\psi}_{\frac{3N}{4}}.\nonumber 
\end{eqnarray}
Then, focusing on the amplitude for the site where the quantum walk
initiated, we obtain after some algebra, 
\begin{equation}
\overline{\psi}_{0}=\left[\mathbb{I}-\left(A_{k}+B_{k}+M_{k}\right)\right]^{-1}\psi_{IC}.\label{eq:psi0in1dsearch}
\end{equation}
Since $A_{k}+B_{k}+M_{k}$ is a unitary matrix for $\left|z\right|=1$,
with an orthonormal set of eigenfunctions and eigenvalues as a function
of $z$, we can expand the matrix on the right-hand side of Eq.\ (\ref{eq:psi0in1dsearch})
in those eigenfunctions and obtain the poles of $\overline{\psi}_{0}$
whenever one (or more) of its eigenvalues become $=1$ for some value
of $z$. We have obtained expressions equivalent to Eq.\ (\ref{eq:psi0in1dsearch})
for other fractal networks, however, these are lengthy and it is cumbersome
to establish the uni-modularity of their poles case-by-case, although
it is always ensured by Eq.\ (\ref{eq:rho_z}).
\end{document}